\documentclass[aps,prb,twocolumn,reprint,superscriptaddress,floatfix]{revtex4-2}
\usepackage{siunitx}
\usepackage{physics}  
\usepackage{graphicx} 
\usepackage{amssymb}
\usepackage{amsmath} 
\usepackage{soul}  
\usepackage{hyperref}
\usepackage{array}

\newcommand{\PdTe}{PdTe$_2$ }
\newcommand{\Hc}{$H_{c2||}$}

\usepackage{multirow}
\usepackage{threeparttable}

\usepackage{array}
\usepackage[table]{xcolor}
\usepackage{threeparttable}
\usepackage{graphicx}   

\begin{document}
\title{g-Factor–Enhanced Upper Critical Field in Superconducting \PdTe due to Quantum Confinement}
\author{Kota Yoshimura}
\affiliation{Department of Physics and Astronomy, University of Notre Dame, Notre Dame, Indiana 46556, USA}
\author{Tzu-Chi Hsieh}
\affiliation{Department of Physics and Astronomy, University of Notre Dame, Notre Dame, Indiana 46556, USA}
\author{Huiyang Ma}
\affiliation{Department of Physics, Florida State University, Tallahassee, Florida, 32306, USA}
\affiliation{National High Magnetic Field Laboratory, 1800 E Paul Dirac Dr, Tallahassee, FL 32310, USA}
\author{Dmitry V. Chichinadze}
\affiliation{National High Magnetic Field Laboratory, 1800 E Paul Dirac Dr, Tallahassee, FL 32310, USA}

\author{Shan Zou}
\affiliation{Department of Physics and Astronomy, University of Notre Dame, Notre Dame, Indiana 46556, USA}
\author{Michael Stuckert}
\affiliation{Department of Physics and Astronomy, University of Notre Dame, Notre Dame, Indiana 46556, USA}
\author{David Graf}
\author{Robert Nowell}
\affiliation{National High Magnetic Field Laboratory, 1800 E Paul Dirac Dr, Tallahassee, FL 32310, USA}
\author{Muhsin Abdul Karim}
\affiliation{Department of Physics and Astronomy, University of Notre Dame, Notre Dame, Indiana 46556, USA}
\author{Daichi Kozawa}
\author{Ryo Kitaura}
\affiliation{Research Center for Materials Nanoarchitectonics (MANA), National Institute for Materials Science (NIMS), 1-1 Namiki, Tsukuba 305-0044, Japan}
\author{Xiaolong Liu}
\author{Xinyu Liu}
\author{Dafei Jin}
\affiliation{Department of Physics and Astronomy, University of Notre Dame, Notre Dame, Indiana 46556, USA}
\author{Cyprian Lewandowski}
\affiliation{Department of Physics, Florida State University, Tallahassee, Florida, 32306, USA}
\affiliation{National High Magnetic Field Laboratory, 1800 E Paul Dirac Dr, Tallahassee, FL 32310, USA}
\author{Yi-Ting Hsu}
\author{Badih A. Assaf*}
\affiliation{Department of Physics and Astronomy, University of Notre Dame, Notre Dame, Indiana 46556, USA}
\email{bassaf@nd.edu}
\date{\today}

\begin{abstract}
    The Pauli limiting field of superconductors determines the maximal possible value of magnetic field at which superconductivity remains possible. For weak-coupling superconductors, it is determined by an established relation that can be found by setting the condensation energy equal to the magnetization free energy. The latter is a function of the carrier $g$-factor. Here, we demonstrate in a van der Waals superconductor \PdTe, that quantum confinement can tune the effective $g-$factor causing the Pauli limit to become thickness dependent. We experimentally probe the in-plane upper critical field $H_{c2||}$ of \PdTe at intermediate thicknesses down to $20 mK$. \Hc is enhanced by more than an order of magnitude as the thickness is varied from $50 nm$ down to $19 nm$. We model the temperature and thickness dependent \Hc, revealing that both orbital and spin Zeeman depairing mechanisms impact its value. While the variation of the orbital interaction is expected, our findings reveal how the Zeeman interaction impacts superconductivity in thin films. They aid in the search for mixed and odd pairing superconductivity where an enhancement of \Hc can be occasionally associated with those unconventional pairing symmetries.
\end{abstract}

\maketitle

\textit{Introduction ---}
The superconducting state of a material can be suppressed by increasing its temperature above the critical temperature, by running a sufficiently large current through it, or by applying a magnetic field large enough to induce Cooper pair depairing. In the latter case, the required field strength is collectively set by orbital, spin, and mixed spin–orbit interactions.  In two-dimensional materials below a critical thickness, applying an in-plane magnetic field can suppress the orbital contributions to depairing, leaving spin-dependent effects as the dominant mechanism. 

The upper critical in-plane field is then set by the fundamental Pauli limit, which is reached when the paramagnetic energy is equal to the superconducting condensation energy determined by the Bardeen-Cooper-Schrieffer (BCS) gap $\Delta$ \cite{clogston1962upper, chandra1962}. In the absence of any other depairing mechanism, the knowledge of the effective $g$-factor and the superconducting gap precisely determine the Pauli limiting field $H_P$ for a given material as:
\begin{equation}
\label{eq:pauli_limit}
    \mu_0 H_P = \frac{\sqrt{2} \Delta}{g \mu_B}, 
\end{equation}where $\Delta\approx1.76 k_B T_c0$ for spin-singlet superconductors in the weak-coupling regime, $T_{c0}$ is the critical temperature at zero magnetic field, $\mu_0$, $\mu_B$, $k_B$ are the permeability of vacuum, the Bohr magneton and the Boltzmann constant respectively. When $g=2$, one recovers the well-known relation $H_P \approx 1.84 T/K \times T_c0$. 

\begin{figure}[ht]
  \centering
  \includegraphics[width=0.48\textwidth]{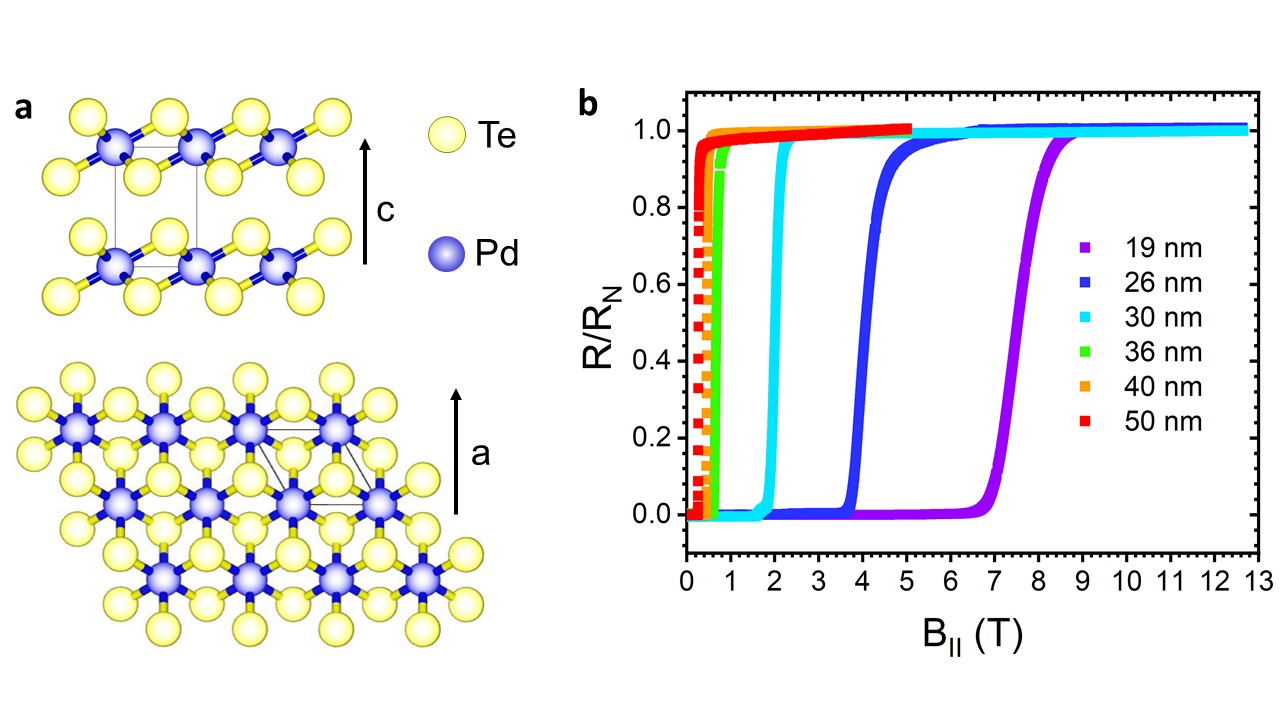}
  \caption{(a) Crystal structure of \PdTe. (b) Normalized resistivity as a function of magnetic field at 400mK for 6 \PdTe films of varying thickness.}
  \label{fig1}
\end{figure}

The violation of the Pauli limit has often been viewed as an indication of an unconventional superconductivity emerging from a normal state with strong spin-orbit coupling \cite{Lu2015, xi2016ising, de2018tuning, AndreaBBGSCScience2022,Zhang2023_Enhanced_SC_proximitized_BBG,Holleis2025_Nemacity_BBG} or superconductivity with a spin-triplet pairing channel\cite{GilAndyMacScience2001,BelitzKirkpatrickPRB2003,AgterbergPRL2004,ShengScience2019,ShengRanNatPhys2019,AokiJPSJ2019,KnebelJPSJ2019,ThomasSciAdv2020,AjeeshPRX2023,CambridgePNAS2024,Cambridge2025PRX_QCL}. Other origins could also include spin-orbit coupling that modifies the normal state's diamagnetic properties and the presence of impurities (Maki theory) \cite{WHHPhysRev1966, FischerHelvetica1973,KlemmLutherBeasley,Fischer1978}. 
However, experimentally, it is difficult to disentangle a violation of the Pauli limit with $g=2$ from 
an enhancement of $H_p$, without knowledge of the effective $g$-factor. The effective $g$-factor of charge carriers is also known to deviate from 2 in various systems that include two-dimensional materials, Dirac semimetals, and conventional narrow-gap semiconductors. For instance, the $g$-factor in MoSe$_2$ was found to be larger than 2 due to the electron-electron interactions \cite{LarentisMoSe2}. In topological materials and semiconductors with quasi-linear band dispersions due to spin-orbit coupling, the effective $g$-factor was also reported to be enhanced by a factor inversely proportional to the energy gap and proportional to the band velocity  \cite{Pidgeon1967, winkler2003spin, guldner1973interband, Landwehr_gfactor, Jiashu_gfactor, bernick1970energy, Suchalkin_InAsSb}. 
In the latter case, quantum confinement that determines the separation between conduction and valence levels is also known to strongly modify $g$ \cite{winkler2003spin}.  

The impact of quantum confinement on the $g-$factor in superconducting thin films, should also manifest as a changing $H_P$ causing a thickness dependent upper critical field as indicated in Eq. \ref{eq:pauli_limit}. This effect is difficult to observe primarily because superconductivity typically requires a large density-of-states to appear at reasonable temperatures and metals with large density-of-states are not as prone to quantum confinement as semiconductors. At the same time, systems where the effective $g$-factor deviates from 2 and is  thickness dependent, usually host charge carriers with small effective mass and have a low density-of-states at the Fermi level ends, too low for superconductivity to emerge. To observe an in-plane upper  critical field that manifests the thickness-dependence in $g$-factor, it is crucial for the superconducting thin films that host both of the following ingredients: (i) some charge carriers with a small effective mass and large $g$ at the Fermi level, and (ii) separate regions of large density-of-states on the Fermi surface.

\begin{figure*}[t]
  \centering
  \includegraphics[width=0.80\textwidth]{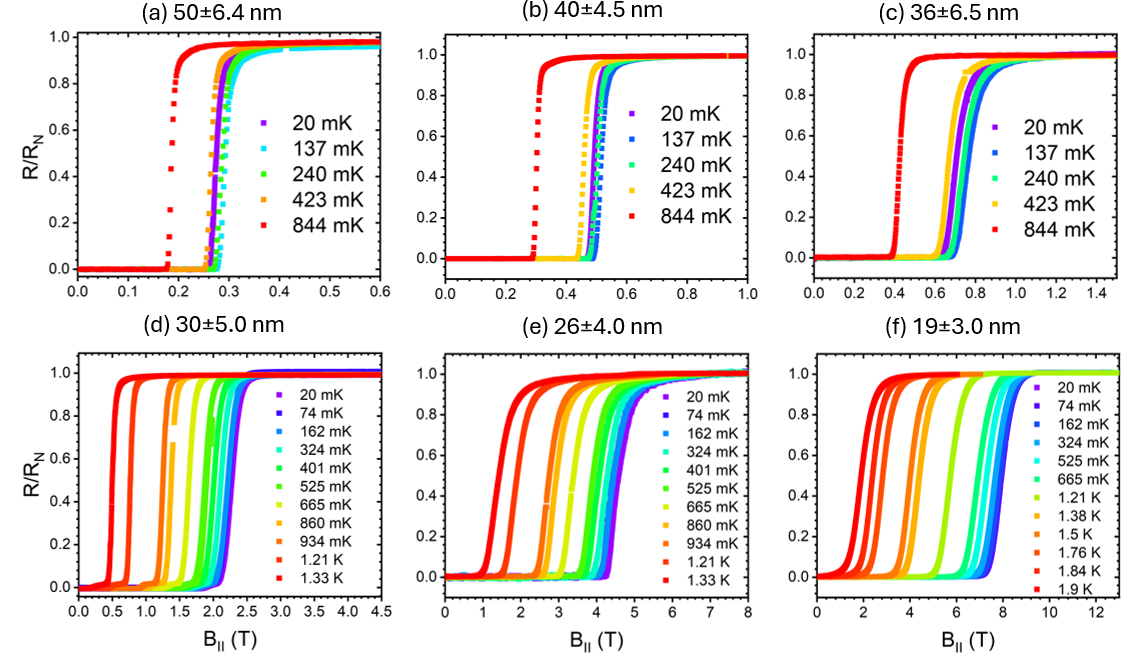}
  \caption{(a)-(f) Magnetic field dependence of Normalized Resistance at different temperatures for a field applied in the sample plane.}
  \label{fig2}
\end{figure*}

We find these two ingredients in the centrosymmetric van der Waals superconductor \PdTe (Fig.~\ref {fig1}(a)). \PdTe hosts a Fermi surface that partially emerges from quasi-linearly dispersing bands, leading to effective masses as small as 0.05$m_0$\cite{chapai2020quantum, amit2018type, liu2018two, leng2017type, kim2018importance} and a significantly enhanced carrier $g$-factor \cite{ZhangBalicas2018}. Below 1.9$K$, \PdTe is known to become a type-II Ising superconductor \cite{leng2017type,liu2018two,liu2020type,das2018conventional}. In this work, in quantum wells of \PdTe with thickness varying between 50nm and 19nm, we report an order of magnitude enhancement of the in-plane \Hc (Fig.~\ref{fig1}(b)) that we relate to a decreasing $g$-factor due to quantum confinement. To capture these experimental trends we implement a self-consistent linearized gap equation description \cite{osti_4763139, Saito2015, Lu2015, AgterbergPRL2004, Zwicknagl2017} and account for the behavior of the in-plane upper critical field $H_{c2||}$  down to 20mK. This approach allows us to reliably extract orbital, Zeeman and spin-orbit coupling energies for each sample. 
Our analysis relates the modulation of \Hc in thin quantum wells, to a strongly thickness-dependent effective $g$-factor. It reveals the impact of quantum confinement on the Pauli limit through a tuning of the Zeeman interaction.
\color{black}

\PdTe thin films are grown by tellurization of Pd layers using a chemical vapor deposition process described in \cite{synthesispdte2, liu2021epitaxial}.  The films are characterized by X-ray diffraction (Suppl. Section 1) which confirms the formation of single phase \PdTe. Atomic force microscopy measurements shown in Suppl. Section 2 allow us to extract the thickness and roughness of each film. The normal state resistivity of six films studied in this work, both at zero field and at a finite field above in the in-plane $H_{c2||}$ are also shown in Suppl. Section 1. An order of magnitude increase in resistivity is seen in the three thinnest samples.

\begin{figure*}[t]
  \centering
  \includegraphics[width=0.80\textwidth]{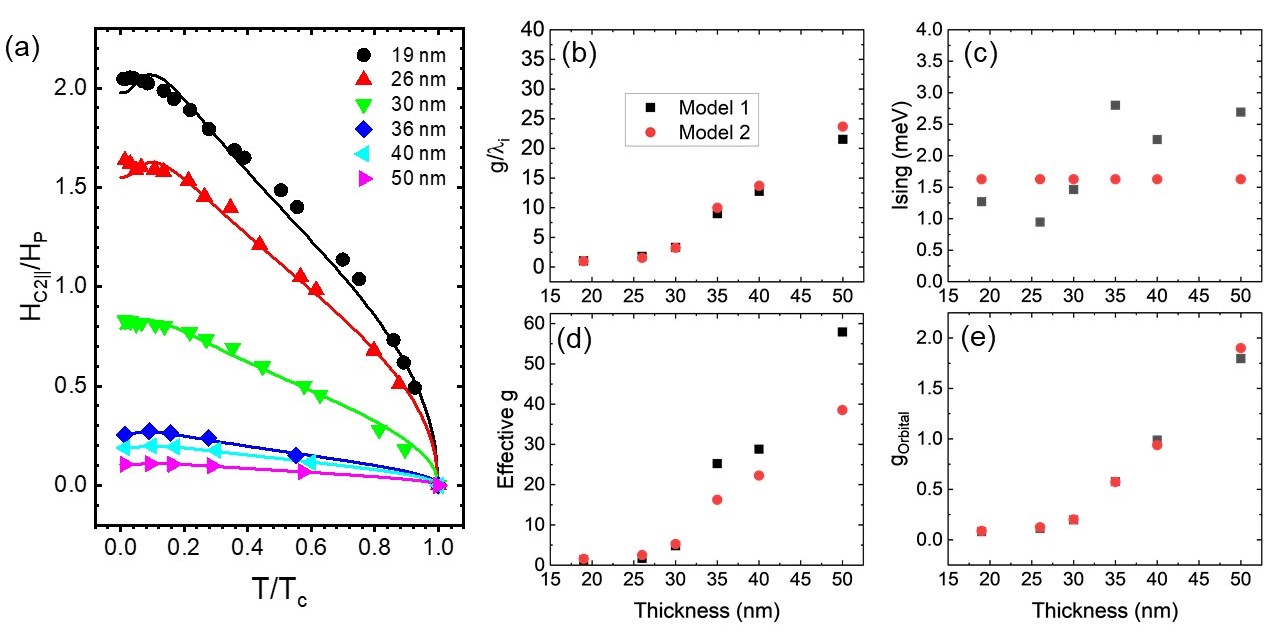}
  \caption{(a) Temperature dependence on the in-plane upper critical field for films of varying thickness. The circles represent experimental data points and the curve fits are from a model where all parameters are varied and the orbital interaction is assumed to be independent on angle. (b) ratio between the effective $g$-factor and the Ising parameter extracted from the fit (c) Ising parameter, (d) effective Zeeman g factor and (e) orbital g. The parameters are shown for two models Model 1 allows 3 parameters (Ising, Zeeman g and orbital g) to vary, Model 2 uses a fixed Ising interaction.}
  \label{fig3}
\end{figure*}

Measurements of the superconducting upper critical fields of \PdTe are carried out down to 20mK at SCM1 at the National High Magnetic Field Lab. Additional measurements are done on-site to characterize the out-of-plane \Hc and $T_{c0}$, above 1.4K. Fig.~\ref{fig2} plots the resistivity normalized by the normal state resistivity  versus magnetic field for the set of six quantum wells. The field is applied in the sample plane, along the current direction. For 50nm \PdTe, $H_{c2||}$  does not exceed 0.1T, when taken as the field at which the resistivity is halved. Comparing the 19nm and 50nm layers to each other, one can notice that $H_{c2||}$ is enhanced by a factor of 27 and reaches 8T for the thinnest sample. $T_{c0}$ does not vary significantly between 50nm and 26nm (1.5K), but is found  to be 2.1K for the 19nm sample (see Suppl. Section 3).

To explore physical implications of the experimental trends, we employ an effective two-band model that is frequently used to study 2D Ising superconductors including PdTe$_2$. \citenum{Saito2015, Lu2015,Liu2020}. Our model, detailed in the SI, includes Ising ($\lambda_I$), Zeeman ($g$) and orbital interactions (${g_{orb}}$) as possible tunable parameters. Note that $\lambda_I$ here is the effective Ising interaction, renormalized by a factor inversely proportional to the scattering time \cite{PhysRevX.8.021002}. We measure the orbital interaction in units of Zeeman energy, thus $g$ and $g_{orb}$ are dimensionless quantities. To determine the relation between in-plane critical field \Hc and critical temperature at that magnetic field ($T_c$), we solve a self-consistent linearized gap equation\cite{osti_4763139, Saito2015, Lu2015,AgterbergPRL2004,Zwicknagl2017}:
\begin{equation}
\begin{gathered}
    \ln \left(\frac{T}{T_{c0}} \right) = \\ T \sum_{\omega} \left[ \left \langle \int \frac{d\epsilon_{\mathbf{k}}}{2} \mathrm{tr} \left\{ \sigma_{y} \hat{G}^{0*}_{\omega} (-\mathbf{k}) \sigma_{y} \hat{G}^{0}_{\omega} (\mathbf{k}) \right \} \right \rangle_{FS} -  \frac{\pi}{|\omega|} \right].
\end{gathered}
\label{Tc_eq}
\end{equation}
In the above equation, the normal state Green's function $G^{0}_{\omega} (\mathbf{k})$ depends on $\lambda_I,g,g_{orb}$. By carrying out a fitting procedure of Eq.\eqref{Tc_eq} to the temperature-dependent \Hc, we can extract the values of Ising,  Zeeman, and orbital couplings. We consider several models accounting for the variations of these coupling to extract their thickness dependence.

Fig. \ref{fig3}(a) plots \Hc as a function of the reduced temperature $T/T_{c0}$ for each of the six samples of varying thickness together with curve fits obtained from Eq.\eqref{Tc_eq}. In the theory fitting shown in Fig. \ref{fig3}(a) we allow for all parameters (Ising $\lambda_I$, Zeeman $g$-factor, and orbital interaction $g_{orb}$) to vary from sample to sample (i.e. allowing for them to be thickness dependent) - we refer to this as Model 1. In our model, we set the Rashba SOC to zero as expected for type-II centrosymmetric Ising superconductors that preserve inversion symmetry. If Rashba SOC is permitted to be non-zero, we found that the best fit is achieved for vanishingly small Rashba SOC (in agreement with earlier experiments in type-II Ising superconductors \cite{PhysRevLett.123.126402,Liu2020}). We find that since the scale of the Ising coupling exceeds $T_{c0}$ significantly, our fitting routine cannot independently constrain both $\lambda_I$ and $g$. Rather we can only constrain their ratio. In the second model we consider, Model 2, we keep $\lambda_I$ fixed across all samples (i.e. Ising is thickness independent), and vary the Zeeman $g$ and $g_{orb}$.  For both models, we test an orbital interaction that is either dependent or independent on Bloch momentum angle as defined in Suppl. Section 4 and we find that the angle independent model yields better fit. Fig. \ref{fig3} (b-e) plot the thickness dependence of our fit parameters obtained from Model 1 and Model 2. Despite the fact that we cannot precisely determine $\lambda_I$, we consistently find both $g_{orb}$ and $g$ increase with increasing thickness. Additional tests including attempts to fit by fixing the value of $g$ shown in Suppl. Section 4 confirm this finding.

Motivated by our observations that the effective $g$-factor in \PdTe grows with thickness, we now investigate the physical origins of this behavior. An enhanced $g$-factor is unexpected in most conventional superconductors and, to the best of our knowledge, has not been considered in past work on van der Waals superconductors. Such thickness-dependence in  $g$-factor is, however, common in narrow-gap semiconductors and topological semimetals. 
According to Roth's formula for semiconductors \cite{Roth1959theory, winkler2003spin}, strong spin-orbit coupling and a small band gap are key ingredients for a large effective $g$-factor both of which are present in \PdTe. 

Experimentally, 
prior quantum oscillation measurements on \PdTe single crystals have probed its Fermi surface and have evidenced the presence of carrier pockets, two among them labeled $\alpha$ and $\beta$ hosting charge carriers with effective masses as low as $m_{\alpha}=0.06m_0$ \cite{chapai2020quantum}. The $\alpha$-band was also found to have carriers Fermi wavevector $k_f$ as small as 0.015\r{A} justifying the fact that quantum confinement should impact the dispersion in films with thicknesses comparable with the Fermi wavelength. There is also evidence of a strongly enhanced effective $g$-factor in \PdTe from quantum oscillations reported in \cite{ZhangBalicas2018}.
These findings reveal the impact of strong-spin orbit coupling on the band structure, justifying the presence of significant interband contributions to the effective mass and the $g$-factor. 

Theoretically, we now explain how 
the effective $g$-factor can grow with thickness in \PdTe thin films, as it does in narrow-gap semiconductors \cite{winkler2003spin}. We perturbatively calculate the effective $g$-factor for a $\mathbf{k} \cdot \mathbf{p}$ model of  \PdTe, where $g$ deviates from 2 due to the coupling to higher-energy bands. We show that this correction to the effective $g$-factor due to higher bands is 
sensitive to the sample thickness due to the resulting subbands under quantum confinement. 
We will walk through the calculation in the following, and show detailed calculations in Suppl. Section 5.

\begin{figure}[ht]
    \centering
    \includegraphics[width=0.95\linewidth]{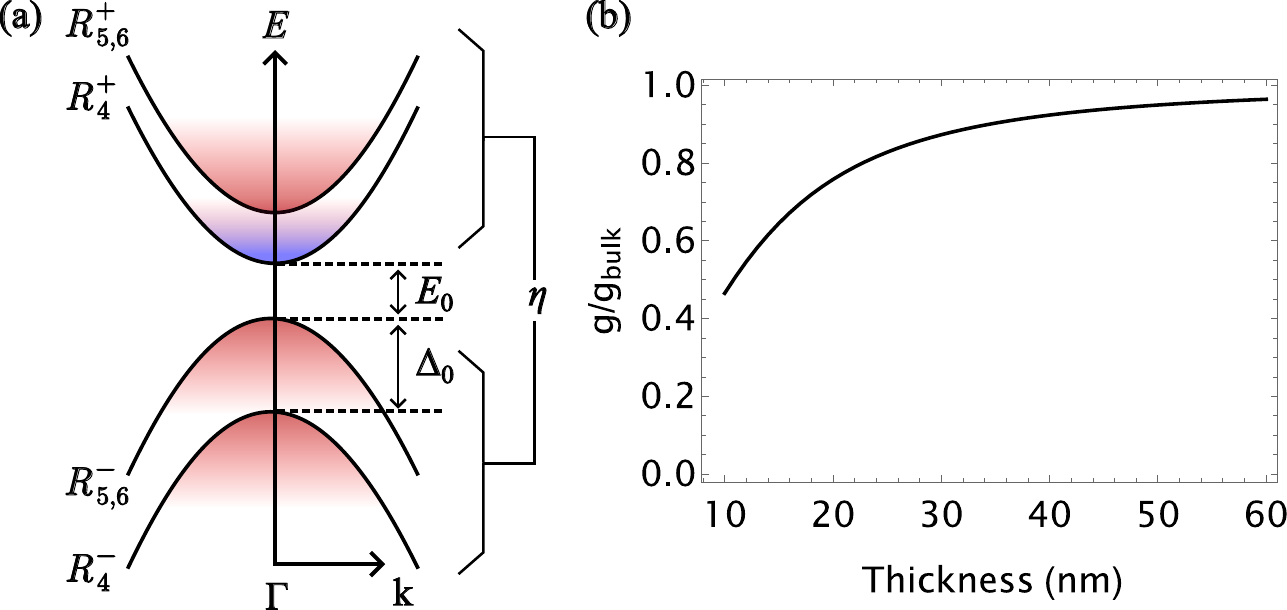}
    \caption{(a) Schematic band structure of PdTe$_2$ near the $\Gamma$ point. The four doubly-degenerate bands are labeled as $R^\pm_4$ ($|j = 3/2,\ |m_j| = 1/2\rangle$) and $R^\pm_{5,6}$ ($|j = 3/2,\ |m_j| = 3/2\rangle$), where the superscript indicates even ($+$) or odd ($-$) parity at the $\Gamma$ point. Here, $j$ and $m_j$ refer to the total angular momentum and its projection. The red shaded regions represent the $p_{x \pm i y}$ character of the bands, while the blue shading in the $R^+_4$ band highlights hybridization with the $p_z$ orbitals. (b) Thickness dependence of the effective $g$-factor (normalized by its bulk value) for the first subband in the valence bands of thin-film PdTe$_2$. The bulk band parameters are set to $E_0 = 0.2\ \mathrm{eV}$, $\Delta_0 = 1\ \mathrm{eV}$, $\eta = 10\ \mathrm{eV}^{-1}$, and effective mass $m = 0.06m_0$.}
    \label{fig:kp_band}
\end{figure}

Consider a $\mathbf{k} \cdot \mathbf{p}$ Hamiltonian $H = H_0 + H_{\boldsymbol{k} \cdot \boldsymbol{p}}$, which consists of four doubly-degenerate bands near the Fermi energy (located between the $R^-_4$ and $R^-_{5,6}$ bands) around the $\Gamma$ point, formed by the product of $p_{x \pm i y}$ orbital, parity, and spin subspaces, as shown in Fig.~\ref{fig:kp_band}(a). The unperturbed Hamiltonian $H_0$ is diagonal with decoupled even and odd parity sectors, while the perturbation $H_{\boldsymbol{k} \cdot \boldsymbol{p}}$ introduces the interband $\mathbf{k} \cdot \mathbf{p}$ couplings for them. Quasi-degenerate perturbation theory yields corrections to the odd-parity bands, leading to an in-plane effective $g$-factor of the form $g = g_0 + \eta \left( \frac{1}{E_0} + \frac{1}{E_0 + \Delta_0} \right)$. The deviation from the bare value $g_0=2$ arises from a finite coefficient $\eta$, which reflects strong interband interactions as well as spin-orbit coupling that hybridizes the $p_z$ and $p_{x \pm i y}$ orbitals of tellurium at avoided crossings. The Ising coupling $\Delta_0$, however, is not essential. In thin films, quantum confinement increases the energy gap $E_0$ between the two parity sectors, and the effective $g$-factor is generally reduced in thinner films. Fig.~\ref{fig:kp_band}(b) shows the thickness dependence of the $g$-factor for the first subband, calculated under the assumptions of infinite well confinement and the effective mass approximation. In real materials, however, both $\eta$ and $E_0$ depend on additional microscopic details. The thickness dependence of the orbital interaction is also further discussed in Suppl. Section 6 as its role in depairing is smaller than that of the Zeeman mechanism.

\begin{figure}[h]
  \centering
  \includegraphics[width=0.48\textwidth]{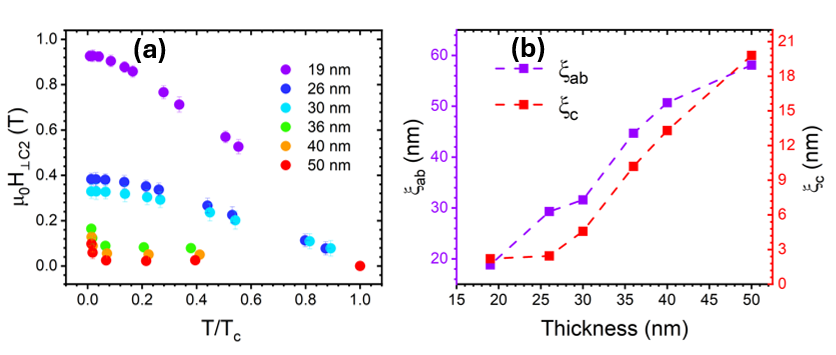}
  \caption{(a) Critical field in out of plane direction. (b) $\xi_{ab}$, $\xi_{c}$ are coherence length in the ab-plane and c-plane direction. Coherence length is calculated using the anisotropic GL model.}
  \label{fig4}
\end{figure}

We also measured the thickness and temperature dependence of the out-of-plane critical field $H_{c2\perp}$ (see Fig. \ref{fig4}(a)) and find that $H_{c2\perp}$ grows with a decreasing sample thickness $d$. It reaches a value significantly smaller than \Hc, expected when orbital interactions are present.

An analysis of Fig. \ref{fig4}(a) still allows us to obtain the coherence length $\xi_{ab}$ and $\xi_{c}$ where $ab$ and $c$ denote the in-plane and out-of-plane directions, respectively. We use the anisotropic Ginzburg-Landau model $ H_{c2\perp}=\frac{\Phi_{0}}{2\pi\xi^{2}_{ab}}$ to derive the in-plane coherence length $\xi_{ab}$. To obtain the out of plane coherence length $\xi_{c}$, we use the relation $\gamma=\frac{H_{c||}}{H_{c\perp}}=\frac{\xi_{ab}}{\xi_{c}}$. Fig. \ref{fig4}(b) shows the thickness dependence both $\xi_{c}$ and $\xi_{ab}$. $\xi_{c}$ comes out consistently smaller than the sample thickness, indicating the films studied here are in a quasi-2D regime for which $d \approx \xi_{ab}$. These findings also support that the enhancement of the critical field is not attributed to the change in thickness with respect to the coherence length.

 Lastly, we note that for the three thinnest films, our model fit in Fig. \ref{fig3} could not explain the lowest temperature region below 100mK. With our experimentally determined $g-$factor, we also still firmly find that all films except 50nm break the adjusted Pauli limit. For the 50nm, the uncertainty on the value of $g$ is too large to confirm a violation. Thus, it is likely that a potentially unconventional pairing mechanism emerges in films but may be absent in bulk \PdTe.

In conclusion, we have observed that the changing effective $g$-factor of \PdTe under quantum confinement results in a dramatic enhancement of its in-plane upper critical field by more than one order of magnitude. Our findings highlight the important role of understanding variations of $g$ with thickness in 2D materials with strong spin-orbit coupling, as a route to tune the Pauli limiting field. It aids in future studies seeking to identify triplet pairing channels in a variety of superconductors.

\textit{Acknowledgment} 
 The Notre Dame group (K.Y., T.C.H, X.L., X.L., Y.T.H. and B.A.A). acknowledge Department of Energy Basic Energy Science Award DE-SC0024291 as the primary source of support for the work (sample synthesis and characterization, measurement of Hc and its analysis, modelling the thickness dependence of the effective $g$-factor). M.A.K. and M.S. who assisted with magnetotransport measurements at low temperature are supported by NSF-DMR-2313441. D.V.C. (modelling Hc versus temperature and the thickness dependence of the orbital interaction) acknowledges financial support from the National High Magnetic Field Laboratory through a Dirac Fellowship. H.M. and C.L. (modelling \Hc versus temperature and the thickness dependence of the orbital interaction) are supported by start-up funds from Florida State University and the National High Magnetic Field Laboratory. A portion of this work was performed at the National High Magnetic Field Laboratory, which is supported by National Science Foundation Cooperative Agreements No. DMR-1644779, No. DMR-2128556, and the State of Florida. D. J. and S. Z. acknowledge support from the Department of Energy (DOE) under Award No. DE-SC0025542 for the development of millikelvin magnetotransport infrastructure at Notre Dame. The work at NIMS (growth recipe development) was supported by JSPS KAKENHI Grants No. 25K22210, 24H01210, 24H02218, 23H05469, JST SCICORP Grant No. JPMJSC2110 and World Premier International Research Center Initiative (WPI), MEXT, Japan.

\bibliographystyle{apsrev4-2}
\bibliography{bib/ref}  

\end{document}